\documentclass[11pt]{article}
\usepackage{amsfonts}
\usepackage{amsmath}
\usepackage[textwidth=160mm, textheight=230mm]{geometry}
\begin{document}
\begin{titlepage}
\begin{flushright}
CP3-07-23\\
August 2007\\
\end{flushright}
\begin{centering}

{\ }\vspace{1cm}

{\Large\bf An Action for Chan-Paton Factors}

\vspace{1.5cm}

Florian Payen

\vspace{1.3cm}

{\em Center for Particle Physics and Phenomenology (CP3)}\\
{\em Institute of Nuclear Physics}\\
{\em Department of Physics, Universit\'e catholique de Louvain}\\
{\em Chemin du Cyclotron 2, B-1348 Louvain-la-Neuve, Belgium}\\
{\em E-mail: {\tt florian\_payen@hotmail.com}}

\vspace{1.0cm}

\begin{abstract}
\noindent We show that Chan-Paton factors can be derived from a
classical action describing the dynamics of a new group-valued
degree of freedom attached to the boundary of an open bosonic
string. We discuss the free and the interacting string in the
oriented and unoriented cases, as well as the coupling of the string
to an external Yang-Mills gauge field, and recover by this approach
well-known results.

\end{abstract}

\vspace{10pt}

\end{centering}

\vspace{25pt}

\end{titlepage}

\setcounter{footnote}{0}

\section{Introduction}

Chan-Paton factors \cite{CP} are matrices $\lambda_{ij}$ associated
with the quantum states of an open bosonic string, in order to
provide them with an internal non-abelian symmetry. Indeed, in a
string interaction process, the factors corresponding to the
external states are collected into traces which are inserted in the
amplitude, leading to non-abelian gauge interactions \cite{GSW,P}.
The $i$ and $j$ indices of the Chan-Paton factors are usually
interpreted as quantum numbers attached to both ends of the string,
and propagating freely during the interaction. In this paper, we
want to show that this picture could be implemented in a natural way
by deriving the Chan-Paton construction from the dynamics of some
new degree of freedom living on the boundary of the string
world-sheet.

A few attempts to put degrees of freedom on the edges of a string
can be found in the literature. In particular, Marcus and Sagnotti
\cite{MS2} proposed successfully to quantize $d$ free fermions on
the string boundary, recovering Chan-Paton factors for the
symmetry group $SO(2^{d/2})$. But in their approach the group
structure only becomes apparent at the quantum level.

Here we chose a different approach, starting from a classical action
introduced by Balachandran, Borchardt and Stern \cite{BBS} to
describe the interaction of a pointlike non-abelian charge with a
Yang-Mills gauge field. The canonical quantization of this system
gives rise to a finite dimensional Hilbert space carrying an
irreducible unitary representation of the gauge group
\cite{BBS,JNP}, while the related path integral provides a
representation for the Wilson loop of the gauge field \cite{DP,B}.
Later, Barbashov and Koshkarov \cite{BK} coupled this action to the
boundary of an open bosonic string in order to describe its
classical interaction with a Yang-Mills gauge field. The purpose of
this paper is to show that at the quantum level this coupling leads
to Chan-Paton factors.

The presentation will be organized as follows. First we will
present the classical action for the new `Chan-Paton' degree of
freedom attached to the boundary of an oriented open bosonic
string. Next we will show that the quantum dynamics of this degree
of freedom leads to the usual Chan-Paton factors, both in the
canonical quantization of a free string and in the path integral
description of string interactions. Then we will extend our
analysis to the case of an unoriented open bosonic string. Finally
we will discuss the interaction of a string with an external
Yang-Mills gauge field.

\section{Action}

Let us consider a Minkowski spacetime $M^D$ with dimension $D$
fixed to the critical value 26, described by the canonical
coordinates $x^{\mu}$ $(\mu = 0,1,\ldots,D-1)$ and the metric
\begin{equation}
\eta_{\mu\nu}=diag(-1,1,\ldots,1).
\end{equation}
We choose units such that $c=1$.

An oriented open bosonic string in $M^D$ is characterized by an
oriented world-sheet $\Sigma$, parameterized by local coordinates
$\sigma^\alpha$ $(\alpha=0,1)$, which admits a boundary
$\partial\Sigma$ with the induced orientation, described by some
function $\bar{\sigma}^{\alpha}(\tau)$ of a parameter $\tau$. The
world-sheet is provided with an intrinsic metric
$\gamma_{\alpha\beta}(\sigma)$ of signature $(-,+)$, and embedded
by coordinates $x^{\mu}(\sigma)$ in $M^D$. The dynamics of the
string is determined by the Polyakov action
\begin{equation}
S_\Sigma[\gamma,x]=-\frac{1}{4\pi\alpha'}\int_\Sigma d^2\sigma\
\sqrt{-\det\gamma}\
\gamma^{\alpha\beta}\eta_{\mu\nu}\partial_{\alpha}x^{\mu}\partial_{\beta}x^{\nu},
\end{equation}
where $\alpha'$ is the Regge slope. It is invariant under Poincar\'e
transformations of the spacetime coordinates, orientation preserving
world-sheet diffeomorphisms and Weyl transformations of the metric.

Now let $G$ be a compact connected Lie group with dimension $N_G$,
$\mathcal{G}$ the Lie algebra of $G$, and $\mathcal{H}$ a Cartan
subalgebra in $\mathcal{G}$. We attach a new `Chan-Paton' degree
of freedom $g(\tau)$ in $G$ to the boundary $\partial\Sigma$ of
the string, whose dynamics is controlled by a new term added to
the action \cite{BBS,BK}
\begin{equation}
S_\Sigma[\gamma,x, g] = S_\Sigma[\gamma,x]+\int_{\partial \Sigma}
d\tau\ \kappa\big(-ig^{-1}\partial_{\tau}g\big).
\end{equation}
Here $\kappa(.)$ is an element of the dual $\mathcal{G}^*$ of
$\mathcal{G}$. Thanks to the right action of $G$ on $g(\tau)$, it
can be univocally reduced to an element of the dual $\mathcal{H}^*$
of $\mathcal{H}$ belonging to the positive Weyl chamber defined by a
family of simple roots. The boundary term preserves the symmetries
of the action, and is invariant
\begin{equation}
S_\Sigma[\gamma,x, g^U]=S_\Sigma[\gamma,x, g]
\end{equation}
under the left action of $G$ on $g(\tau)$
\begin{equation}
g(\tau)\ \rightarrow\ g^U(\tau)=Ug(\tau).
\end{equation}

\section{The Free String}

Let us first consider a free oriented open bosonic string,
characterized by a topologically trivial world-sheet parametrized
by $(\tau,\sigma)$, with boundaries described by $(\tau,0)$ and
$(\tau,\pi)$, and let us study the dynamics of Chan-Paton degrees
of freedom $g_0(\tau)$ and $g_\pi(\tau)$ attached to its ends
\begin{equation}
S[\gamma,x,g_0,g_\pi] = S[\gamma,x]+\int d\tau\
\kappa\big(-ig_0{}^{-1}\partial_{\tau}g_0\big)-\int d\tau\
\kappa\big(-ig_\pi{}^{-1}\partial_{\tau}g_\pi\big).
\end{equation}

Let $(T_a)_{a=1,\ldots,N_G}$ be a basis in $\mathcal{G}$,
satisfying the algebra $[T_a,T_b]=if_{ab}{}^{c}T_c$, where the
$f_{ab}{}^c$ are structure constants. The classical equations of
motion for $g_0(\tau)$ and $g_\pi(\tau)$ \cite{BBS,BK}
\begin{equation}
\partial_\tau G_{0a}(\tau)=0, \qquad \partial_\tau G_{\pi a}(\tau)=0
\end{equation}
express the conservation of the Noether charges
\begin{equation}
G_{0a}(\tau) = - \kappa\big(g_0{}^{-1}(\tau)T_a g_0(\tau)\big),
\qquad G_{\pi a}(\tau) = \kappa\big(g_\pi{}^{-1}(\tau)T_a
g_\pi(\tau)\big)
\end{equation}
associated with the left action of $G$ on the degrees of freedom
\begin{equation}
g_0(\tau)\ \rightarrow\ g_0{}^U(\tau)=U g_0(\tau), \qquad
g_\pi(\tau)\ \rightarrow\ g_\pi{}^U(\tau)=U g_\pi(\tau).
\end{equation}
At the quantum level, choosing units such that $\hbar=1$, the
corresponding operators $\hat{G}_{0a}$ and $\hat{G}_{\pi a}$
satisfy the algebra
\begin{equation}
[\hat{G}_{0a},\hat{G}_{0b}]=if_{ab}{}^c\hat{G}_{0c}, \qquad
[\hat{G}_{\pi a},\hat{G}_{\pi b}]=if_{ab}{}^c\hat{G}_{\pi c},
\end{equation}
and generate some representation of $G$ on the Hilbert space of
quantum states.

As stated in \cite{BBS,JNP}, the canonical quantization of the
system restricts the possible values of $\kappa(.)$. Indeed, the
sector associated with $g_0(\tau)$ can be quantized if and only if
$\kappa(.)$ corresponds to the highest weight of an irreducible
unitary representation $R$ of $G$ on a vector space $V$ with
dimension $N_R$. Then the sector associated with $g_\pi(\tau)$ can
be quantized similarly, as $-\kappa(.)$ corresponds to the lowest
weight of the dual representation $R^*$ of $G$ on the dual space
$V^*$, conjugated to its highest weight by some element of the
Weyl group of $\mathcal{H}$. If this condition is satisfied, the
Hilbert spaces of the two sectors are finite dimensional and
coincide with $V$ and $V^*$, while the operators $\hat{G}_{0a}$
and $\hat{G}_{\pi a}$ are identified with the generators of the
representations $R$ and $R^*$.

Thus the quantum states of the string coupled to Chan-Paton
degrees of freedom take the form $\lambda_{ij}\,|X\rangle$, where
$\lambda_{ij}$ $(i,j=1,\ldots,N_R)$ is a tensor in $V\otimes V^*$
and $|X\rangle$ is a naked string state. The operators
$\hat{G}_{0a}$ and $\hat{G}_{0a}$ act on these states following
\begin{eqnarray}
\hat{G}_{0a}\lambda_{ij}\,|X\rangle
& = & R(T_a)_{ii'}\lambda_{i'j}\,|X\rangle\\
& = & (R(T_a)\lambda)_{ij}\,|X\rangle,\nonumber
\end{eqnarray}
\begin{eqnarray}
\hat{G}_{\pi a}\lambda_{ij}\,|X\rangle
& = & -R^*(T_a)_{jj'}\lambda_{ij'}\,|X\rangle\\
& = & (-\lambda R(T_a))_{ij}\,|X\rangle,\nonumber
\end{eqnarray}
and generate their transformations under the left action of $G$
\begin{eqnarray}
\lambda_{ij}\,|X\rangle \ \rightarrow \ \lambda_{ij}{}^U\,|X\rangle
& = & R(U)_{ii'}R^*(U)_{jj'}\lambda_{i'j'}\,|X\rangle\\
& = &(R(U)\lambda R^\dagger(U))_{ij}\,|X\rangle,\nonumber
\end{eqnarray}
where the operators $R(U)$ in $V$ form a subgroup of $U(N_R)$.

We finally recover the usual Chan-Paton factors $\lambda_{ij}$
associated with the quantum states of an oriented open bosonic
string \cite{GSW,P}.

\section{The Interacting String}

Let us now consider the interactions of oriented open bosonic
strings, and study the dynamics of Chan-Paton degrees of freedom
attached to their ends.

Let $v_\kappa$ be the highest weight vector of the representation
$R$ in $V$, satisfying
\begin{equation}
R(h)\, v_\kappa = \kappa(h)\, v_\kappa \quad \forall
h\in\mathcal{H}, \qquad v_\kappa{}^\dagger v_\kappa = 1.
\end{equation}
On the one hand, it allows us to rewrite the action, in its
euclidean form, as \cite{B}
\begin{eqnarray} S_\Sigma{}^E[\gamma,x, g] & = &
S_\Sigma{}^E[\gamma,x]-i\int_{\partial
\Sigma} d\tau\ \kappa\big(-ig^{-1}\partial_{\tau}g\big)\\
& = & S_\Sigma{}^E[\gamma,x]-i\int_{\partial \Sigma} d\tau\
\big(v_\kappa{}^\dagger R(-ig^{-1}\partial_{\tau}g) v_\kappa\big).
\end{eqnarray}
On the other hand, it enables us to build from the vertex
functional associated with the naked state $|X\rangle$
\begin{equation}
V_\Sigma{}^{(X)}[\gamma,x]= \int_{\partial \Sigma} d\tau\
\mathcal{V}^{(X)}[\gamma,x]
\end{equation}
a new vertex functional associated with the state
$\lambda_{ij}\,|X\rangle$
\begin{equation}
V_\Sigma{}^{(\lambda X)}[\gamma,x,g]= \int_{\partial \Sigma}
d\tau\ \big(v_\kappa{}^\dagger R^\dagger(g) \lambda R(g)
v_\kappa\big)\ \mathcal{V}^{(X)}[\gamma,x].
\end{equation}
The new vertex functional remains covariant under Poincar\'e
transformations of the spacetime coordinates, and stays invariant
under orientation preserving world-sheet diffeomorphisms and Weyl
transformations of the metric. It is also covariant
\begin{equation}
V_\Sigma{}^{(\lambda^U X)}[\gamma,x,g^U]=V_\Sigma{}^{(\lambda
X)}[\gamma,x,g]
\end{equation}
under the left action of $G$ on $g(\tau)$
\begin{equation}
g(\tau) \ \rightarrow \ g^U(\tau)=Ug(\tau), \qquad
\lambda_{ij}\,|X\rangle \ \rightarrow \ \lambda_{ij}{}^U\,|X\rangle
=(R(U)\lambda R^\dagger(U))_{ij}\,|X\rangle.
\end{equation}

Now we can define the $\mathrm{S}$-matrix element describing the
interactions of $S$ external states $\lambda^s{}_{ij}\,|X^s\rangle$
as the euclidean path integral
\begin{equation}
\mathrm{S}\big(\lambda^1X^1,\ \ldots,\ \lambda^S
X^S\big)=i\sum_\Sigma\int[\mathcal{D}\gamma]\int[\mathcal{D}x]\int[\mathcal{D}g]\
\mathrm{g}^{-\mathcal{X}_\Sigma}\ e^{-S_\Sigma{}^E[\gamma,x,g]}\
\prod_{s=1}^S V_\Sigma{}^{(\lambda^sX^s)}[\gamma,x,g].
\end{equation}
The product of the vertex functionals $V_\Sigma{}^{(\lambda^s
X^s)}[\gamma,x,g]$ associated with the external states is weighted
by the exponential of the euclidean action $S_\Sigma{}^E[\gamma,x,
g]$ and a power $\mathrm{g}^{-\mathcal{X}_\Sigma}$ of the string
coupling, where $\mathcal{X}_\Sigma$ denotes the Euler
characteristics of the world-sheet. Then the result is summed over
all compact oriented world-sheets $\Sigma$, euclidean metrics
$\gamma_{\alpha\beta}(\sigma)$, space-time embeddings
$x^\mu(\sigma)$ and Chan-Paton degrees of freedom $g(\tau)$,
inequivalent under orientation preserving world-sheet
diffeomorphisms and Weyl transformations of the metric.

In order to give a meaning to the path integral over the
Chan-Paton degrees of freedom
\begin{equation}
\int[\mathcal{D}g]\ e^{i\int_{\partial \Sigma} d\tau\
\big(v_\kappa{}^\dagger R(-ig^{-1}(\tau)\partial_{\tau}g(\tau))
v_\kappa\big)}\ \prod_{s=1}^S \big(v_\kappa{}^\dagger
R^\dagger(g(\tau^s)) \lambda^s R(g(\tau^s)) v_\kappa\big),
\end{equation}
we choose discrete values $\tau_n$ of the continuous parameter
$\tau$ describing the boundary of the string, some of which,
$\tau_{n^s}$, coincide with the insertion points $\tau^s$ of the
vertex functionals. Following \cite{B}, we then define the path
integral as the continuum limit of the expression
\begin{equation}
\prod_n\Big(N_R \int \mathcal{D}g_n\Big)\ \prod_n
\Big(1+i(\tau_{n+1}-\tau_n) \big(v_\kappa{}^\dagger
R(-ig_n{}^{-1}\frac{g_{n+1}-g_n}{\tau_{n+1}-\tau_n})
v_\kappa\big)\Big)
\end{equation}
\begin{equation*}
\times \prod_{s=1}^S \frac{\big(v_\kappa{}^\dagger
R^\dagger(g_{n^s}) \lambda^s R(g_{n^s+1}) v_\kappa\big)}{
\big(v_\kappa{}^\dagger R^\dagger(g_{n^s})R(g_{n^s+1})v_\kappa\big)}
\end{equation*}
\begin{equation}
=\prod_n\Big(N_R \int \mathcal{D}g_n\Big)\  \prod_n
\big(v_\kappa{}^\dagger R^\dagger(g_n)R(g_{n+1})v_\kappa\big)\
\prod_{s=1}^S \frac{\big(v_\kappa{}^\dagger R^\dagger(g_{n^s})
\lambda^s R(g_{n^s+1}) v_\kappa\big)}{ \big(v_\kappa{}^\dagger
R^\dagger(g_{n^s})R(g_{n^s+1})v_\kappa\big)},
\end{equation}
where $\mathcal{D}g$ is the Haar measure on $G$. But the operator
\begin{equation}
N_R \int \mathcal{D}g\ R(g)v_\kappa v_\kappa{}^\dagger
R^\dagger(g),
\end{equation}
which commutes with all operators $R(U)$ and has trace $N_R$,
defines a resolution of unity on $V$. Thus the discretized form of
the path integral, which consists in a product of such operators
with the Chan-Paton factors inserted at some places, reduces
simply to the expression
\begin{equation}
\mathrm{Tr}\ \mathrm{T}_{\partial \Sigma}\ \prod_{s=1}^S
\lambda^s,
\end{equation}
where the time ordering $\mathrm{T}_{\partial \Sigma}$ indicates
that the cyclic order of the Chan-Paton factors in the trace must
follow that of the insertions of vertex functionals along the
boundary of the string.

We finally recover the usual insertion of traces of Chan-Paton
factors in the amplitudes describing the interactions of oriented
open bosonic strings \cite{GSW,P}.

\section{The Unoriented Case}

Let us now extend our analysis to the case of an unoriented open
bosonic string.

Now the world-sheet $\Sigma$ of the string is not oriented, and its
boundary $\partial\Sigma$ can be given an arbitrary orientation. So
any diffeomorphism $\tilde{\sigma}^{\alpha}(\sigma)$ of $\Sigma$,
which must leave the action $S[\gamma,x,g]$ invariant, induces a
diffeomorphism $\tilde{\tau}(\tau)$ of $\partial\Sigma$ satisfying
\begin{equation}
\tilde{\sigma}^\alpha\big(\bar{\sigma}(\tau)\big)=\bar{\sigma}^\alpha\big(\tilde{\tau}(\tau)\big),
\end{equation}
which either preserves $(d\tilde{\tau}/d\tau>0)$ or reverses
$(d\tilde{\tau}/d\tau<0)$ this orientation. In the second case, the
natural transformation of the Chan-Paton degrees of freedom
\begin{equation}
g(\tau) \ \rightarrow \
\tilde{g}(\tilde{\tau})=g(\tau(\tilde{\tau}))
\end{equation}
reverses the sign of the boundary term of the action. To compensate
for this sign change, we must choose $\kappa(.)$ in such a way that
it is conjugated to $-\kappa(.)$ by an element $W$ in $G$
\begin{equation}
\kappa(W^{-1}.\,W)=-\kappa(.),
\end{equation}
which can be taken in the Weyl group of $\mathcal{H}$. Then the
modified transformation
\begin{eqnarray}
g(\tau) \ \rightarrow \ \tilde{g}(\tilde{\tau}) & = &
g(\tau(\tilde{\tau}))\qquad \quad\textrm{if}\
\frac{d\tilde{\tau}}{d\tau}>0,\\
& = & g(\tau(\tilde{\tau}))W\qquad \textrm{if}\
\frac{d\tilde{\tau}}{d\tau}<0,\nonumber
\end{eqnarray}
leaves the boundary term of the action invariant.

In order to quantize the system successfully, we know that
$\kappa(.)$ must correspond to the highest weight of a
representation $R$ of $G$, $-\kappa(.)$ being conjugated to the
highest weight of the dual representation $R^*$. But as $\kappa(.)$
is now conjugated to $-\kappa(.)$, the highest weights of the two
representations coincide, and the representations $R$ and $R^*$ are
unitarily equivalent. This means that there exists a unitary
operator $M$ from $V^*$ to $V$ such that
\begin{equation}
MR^*(T_a)=-R(T_a)M, \qquad MR^*(U)=R(U)M.
\end{equation}
Then $M^*$ is a unitary operator from $V$ to $V^*$ such that
\begin{equation}
M^*R(T_a)=-R^*(T_a)M^*,\qquad M^*R(U)=R^*(U)M^*.
\end{equation}
The operators $M$ and $M^*$ are unique up to a phase, implying the
relations
\begin{equation}
M^{T}=\alpha M, \qquad M^{\dagger}=\alpha M^*,
\end{equation}
with $\alpha=\pm1$. Moreover it is possible to choose a basis of
$V$ such that $M$ reduces to
\begin{equation}
M=I
\end{equation}
in the first case, or to
\begin{equation}
M=\Big(\begin{array}{cc} 0 & I \\ -I & 0 \end{array}\Big)
\end{equation}
in the second case, with $N_R$ even. The representation $R$ is then
said to be real or pseudoreal, and the operators $R(U)$ in $V$ form
a subgroup of $SO(N_R)$ or $USp(N_R)$.

The physical quantum states of a free unoriented open bosonic
string must be invariant under world-sheet parity, which takes
$(\sigma,\tau)$ to $(\pi-\sigma,\tau)$ and thus exchanges the two
ends of the string. The Chan-Paton degrees of freedom transform as
\begin{equation}
\tilde{g}_0(\tau)=g_\pi(\tau)W, \qquad
\tilde{g}_\pi(\tau)=g_0(\tau)W,
\end{equation}
inducing the transformation of the Noether charges
\begin{equation}
\tilde{G}_{0a}(\tau)=G_{\pi a}(\tau), \qquad \tilde{G}_{\pi
a}(\tau)=G_{0a}(\tau).
\end{equation}
At the quantum level, the corresponding unitary operator
$\hat{\Omega}$ must satisfy
\begin{equation}
\hat{\Omega}\hat{G}_{0a}=\hat{G}_{\pi a}\hat{\Omega}, \qquad
\hat{\Omega}\hat{G}_{\pi a}=\hat{G}_{0 a}\hat{\Omega},
\end{equation}
while reproducing the usual transformation of the naked states
$|X\rangle$
\begin{equation}
\hat{\Omega}|X\rangle=(-1)^{N_X}|X\rangle,
\end{equation}
with $N_X$ measuring their level of excitation. It is then
determined up to a phase $\varepsilon$
\begin{eqnarray}
\hat{\Omega} \lambda_{ij}\,|X\rangle & = &
\varepsilon(-1)^{N_X}M_{ij'}M^*{}_{ji'}\lambda_{i'j'}\,|X\rangle\\
& = &
\varepsilon(-1)^{N_X}(M\lambda^TM^\dagger)_{ij}\,|X\rangle.\nonumber
\end{eqnarray}
Thus states are invariant under world-sheet parity
\begin{equation}
\hat{\Omega} \lambda_{ij}\,|X\rangle=\lambda_{ij}\,|X\rangle
\end{equation}
if the associated Chan-Paton factors verify the condition
\begin{equation}
(M\lambda^TM^\dagger)_{ij}=\varepsilon(-1)^{N_X}\lambda_{ij},
\end{equation}
where the phase $\varepsilon$ have to be restricted to $\pm 1$ to
insure their existence.

The vertex functionals associated with the physical quantum states
of an unoriented open bosonic string must be invariant under any
world-sheet diffeomorphism $\tilde{\sigma}^{\alpha}(\sigma)$, in
particular when the corresponding boundary diffeomorphism
$\tilde{\tau}(\tau)$ is orientation reversing
$(d\tilde{\tau}/d\tau<0)$. Recalling the transformation law of the
naked vertex functional
\begin{equation}
V^{(X)}[\tilde{\gamma},\tilde{x}]=(-1)^{N_X}V^{(X)}[\gamma,x],
\end{equation}
we have
\setlength\arraycolsep{2pt}
\begin{eqnarray}
V_\Sigma{}^{(\lambda X)}[\tilde{\gamma},\tilde{x},\tilde{g}] & = &
\int_{\partial \Sigma} d\tilde{\tau}\ \big(v_\kappa{}^\dagger
R^\dagger(\tilde{g}) \lambda R(\tilde{g}) v_\kappa\big)\
\mathcal{V}^{(X)}[\tilde{\gamma},\tilde{x}]\\
& = & (-1)^{N_X} \int_{\partial \Sigma} d\tau\
\big(v_\kappa{}^\dagger R^{\dagger}(W)R^\dagger(g) \lambda
R(g)R(W) v_\kappa\big)\ \mathcal{V}^{(X)}[\gamma,x]\\ & = &
(-1)^{N_X} \int_{\partial \Sigma} d\tau\
\big(v_\kappa{}^{*\dagger} R^{*\dagger}(W)R^{*\dagger}(g)
\lambda^T R^*(g)R^*(W) v_\kappa{}^*\big)\
\mathcal{V}^{(X)}[\gamma,x].
\end{eqnarray}
But the vectors $R^*(W) v_\kappa{}^*$ and $M^\dagger v_\kappa$ in
$V^*$, which satisfy
\begin{equation}
-R^*(h)\,R^*(W)v_\kappa{}^*=\kappa(h)\,R^*(W)v_\kappa{}^*\quad\forall
h\in\mathcal{H}
\end{equation}
and
\begin{equation}
-R^*(h)\,M^\dagger v_\kappa=\kappa(h)\,M^\dagger
v_\kappa\quad\forall h\in\mathcal{H},
\end{equation}
are both highest weights vectors of the representation $R^*$, and
are therefore equal up to a phase. So we have again
\setlength\arraycolsep{2pt}
\begin{eqnarray}
V_\Sigma{}^{(\lambda X)}[\tilde{\gamma},\tilde{x},\tilde{g}] & = &
(-1)^{N_X} \int_{\partial \Sigma} d\tau\ \big(v_\kappa{}^{\dagger}
MR^{*\dagger}(g) \lambda^T R^*(g)M^\dagger
v_\kappa\big)\ \mathcal{V}^{(X)}[\gamma,x]\\
& = & (-1)^{N_X} \int_{\partial \Sigma} d\tau\
\big(v_\kappa{}^{\dagger} R^{\dagger}(g) M\lambda^TM^\dagger R(g)
v_\kappa\big)\ \mathcal{V}^{(X)}[\gamma,x].
\end{eqnarray}
Thus vertex functionals are invariant under any world-sheet
diffeomorphism
\begin{equation}
V_\Sigma{}^{(\lambda
X)}[\tilde{\gamma},\tilde{x},\tilde{g}]=V_\Sigma{}^{(\lambda
X)}[\gamma,x,g]
\end{equation}
if the associated Chan-Paton factors verify the condition
\begin{equation}
(M\lambda^TM^\dagger)_{ij}=(-1)^{N_X}\lambda_{ij},
\end{equation}
where the ambiguous sign $\varepsilon$ is now fixed to 1.

We finally recover the usual constraint on the Chan-Paton factors
associated with the quantum states of an unoriented open bosonic
string \cite{GSW,P}.

\section{String in an External Yang-Mills Gauge Field}

Let us finally take advantage of the Chan-Paton degrees of freedom
attached to the ends of an open bosonic string to couple it to an
external Yang-Mills gauge field $A_\mu(x)$ in $\mathcal{G}$. The
action reads \cite{BBS,BK}
\begin{equation}
S_\Sigma[x,\gamma, g; A] = S_\Sigma[x,\gamma]+\int_{\partial
\Sigma} d\tau\ \kappa\big(-ig^{-1}(\partial_{\tau}+iA_\tau)g\big),
\end{equation}
where
\begin{equation}
A_\tau(\tau)=A_\mu\big(x(\bar{\sigma}(\tau))\big)\,\partial_\alpha
x^\mu(\bar{\sigma}(\tau))\,\partial_\tau\bar{\sigma}^{\alpha}(\tau)
\end{equation}
is the pullback of the gauge field on the boundary of the string.
It is invariant
\begin{equation}
S_\Sigma[x,\gamma, g^U; A^U]=S_\Sigma[x,\gamma, g; A]
\end{equation}
under the gauge transformations
\begin{equation}
A_\mu{}^U(x)=U(x)A_\mu(x)U^{-1}(x)-i U(x)\partial_\mu U^{-1}(x),
\end{equation}
\begin{equation}
g^U(\tau)=U\big(x(\bar{\sigma}(\tau))\big)g(\tau),
\end{equation}
and generalizes the well-known coupling to an abelian gauge field
\cite{ACNY,FT}.

We can define the effective action for an external Yang-Mills gauge
field $A_\mu(x)$ in the background of interacting open bosonic
strings as the euclidean path integral
\begin{equation}
\mathrm{S}[A]=i\sum_\Sigma\int[\mathcal{D}\gamma]\int[\mathcal{D}x]\int[\mathcal{D}g]\
\mathrm{g}^{-\mathcal{X}_\Sigma}\ e^{-S_\Sigma{}^E[\gamma,x,g;A]},
\end{equation}
in which the cancelation of Weyl anomalies gives rise to equations
of motion for the gauge field. Following \cite{B}, we define the
path integral over the Chan-Paton degrees of freedom
\begin{equation}
\int[\mathcal{D}g]\ e^{i\int_{\partial \Sigma}d\tau\
\big(v_\kappa{}^\dagger
R(-ig^{-1}(\tau)(\partial_{\tau}+A_\tau(\tau))g(\tau))
v_\kappa\big)}
\end{equation}
as the continuum limit of the expression
\begin{equation}
\prod_n\Big(N_R \int \mathcal{D}g_n\Big)\ \prod_n
\Big(1+i(\tau_{n+1}-\tau_n) \big(v_\kappa{}^\dagger
R(-ig_n{}^{-1}\frac{g_{n+1}-g_n}{\tau_{n+1}-\tau_n}+g_n{}^{-1}A_ng_{n+1})
v_\kappa\big)\Big)
\end{equation}
\begin{equation}
=\prod_n\Big(N_R \int \mathcal{D}g_n\Big)\  \prod_n
\big(v_\kappa{}^\dagger
R^\dagger(g_n)(1+i(\tau_{n+1}-\tau_n)R(A_n))R(g_{n+1})v_\kappa\big)
\end{equation}
\begin{equation}
=\mathrm{Tr}\ \prod_n (1+i(\tau_{n+1}-\tau_n)R(A_n)).
\end{equation}
It reduces to a discretized form of the trace of the Wilson loop
of the gauge field along the boundary of the string
\begin{equation}
\mathrm{Tr}\ \mathrm{T}_{\partial \Sigma}\
e^{i\int_{\partial\Sigma}d\tau\ R(A_\tau(\tau))}.
\end{equation}

We finally recover the usual definition of the effective action
for an external Yang-Mills gauge field in the background of
interacting open bosonic strings \cite{FT,DO}.

\section{Conclusion}

In this paper, we studied a new group-valued degree of freedom
attached to the boundary of an open bosonic string. First we showed
that taking it into account in the canonical quantization of a free
string gives rise to the famous Chan-Paton factors, while
introducing it in the path integral description of string
interactions leads to the usual insertion of traces of such factors
in the amplitudes. Then we explained how this approach reproduces
the traditional constraint on these factors in the unoriented case.
Finally we indicated how it can be used to describe the coupling of
an open bosonic string to an external Yang-Mills gauge field.

Our approach thus provides a natural way to introduce non-abelian
interactions in open bosonic string theory, starting from a
classical action, and reproducing the Chan-Paton construction at
the quantum level. It must naturally be completed by further
well-known results, such as the necessity to choose the full
symmetry groups $U(N)$, $SO(N)$ or $USp(N)$ in their fundamental
representation to insure unitarity at the tree level, or even more
severe restrictions imposed by quantum consistency at the one-loop
level.

The discussion could be extended in many ways. Attaching
Chan-Paton degrees of freedom to the boundary of type $I$
superstrings should give similar results. The study of their
$T$-dual picture could give a new insight on the interaction of
strings with $D$-branes.

\section*{\bf Acknowledgements}

I thank Professor J. Govaerts for his careful reading of the
manuscript. This work was supported by the National Fund for
Scientific Research (F.N.R.S., Belgium) through an Aspirant Research
Fellowship, and by the Institut Interuniversitaire des Sciences
Nucl\'eaires and the Belgian Federal Office for Scientific,
Technical and Cultural Affairs through the Interuniversity
Attraction Poles (IAP) P6/11.

\end{document}